\documentclass[reprint,aps,pre]{revtex4-1}
\usepackage{graphicx}
\usepackage [novolumeabbr] {unitsdef} 

\newunit{\torr}{Torr}	
\newunit{\nanovolt}{nV} 
\newunit{\microvolt}{\mu V} 
\newunit{\Kelvin}{\degree\kelvin} 
\newunit{\gigapascal}{G\pascal} 
\newunit{\megapascal}{M\pascal} 
\newunit{\ppm}{ppm} 
\newunit{\nanowatt}{n\watt} 
\newunit{\microwatt}{\mu \watt} 

\begin{document}
\title {Forced and self-excited oscillations of an optomechanical cavity} 

\author{Stav~Zaitsev}
\email{e-mail: zzz@tx.technion.ac.il}

\author{Ashok Kumar Pandey}
\thanks{Current address: Department of Mechanical Engineering, Indian Institute of Technology Hyderabad, Andhra Pradesh, India}

\author{Oleg~Shtempluck}
\author{Eyal~Buks}
\affiliation{Department of Electrical Engineering, Technion - Israel Institute of Technology, Haifa 32000, Israel}


\begin{abstract}
We experimentally study forced and self-excited oscillations of an optomechanical cavity which is formed between a fiber Bragg grating that serves as a static mirror and between a freely suspended metallic mechanical resonator that serves as a moving mirror. In the domain of small amplitude mechanical oscillations, we find that the optomechanical coupling is manifested as changes in the effective resonance frequency, damping rate and cubic nonlinearity of the mechanical resonator. Moreover, self-excited oscillations of the micromechanical mirror are observed above a certain optical power threshold. A comparison between the experimental results and a theoretical model that we have recently derived and analyzed yields a good agreement. The comparison also indicates that the dominant optomechanical coupling mechanism is the heating of the metallic mirror due to optical absorption.

\end{abstract}
\maketitle


\section{Introduction}
\label{sec:introduction}

Studies combining mechanical elements in optical resonance cavities \cite{Braginsky&Manukin_67, Hane&Suzuki_96} experience a significant surge in popularity in recent years due to the fast progress made in both microelectromechanical systems (MEMS) and optical microcavities. For example, optomechanical coupling of nanomechanical mirror resonators to optical modes of high-finesse cavities mediated by radiation pressure has a promise of bringing the mechanical resonators into the quantum realm \cite{Kimble_et_al_01, Carmon_et_al_05, Arcizet_et_al_06, Gigan_et_al_06, Jayich_et_al_08, Schliesser_et_al_08, Genes_et_al_08, Kippenberg&Vahala_08, Teufel_et_al_10}. Furthermore, the micro-optoelectromechanical systems (MOEMS) are expected to play an increasing role in optical communications \cite{Wu_et_al_06} and other photonics applications \cite{Stokes_et_al_90, Lyshevski&Lyshevski_03, Hossein-Zadeh&Vahala_10}.

In addition to the radiation pressure, another important force that contributes to the optomechanical coupling in MOEMS is the bolometric force \cite{Metzger&Karral_04, Jourdan_et_al_08,  Metzger_et_al_08, Marino&Marin_10, Ludwig_et_al_07, Restrepo_et_al_10, Liberato_et_al_10}, also known as the thermal force. This force can be attributed to the thermoelastic deformations of the micromechanical mirrors. In general, the thermal force plays an important role in relatively large mirrors, in which the thermal relaxation rate is comparable to the mechanical resonance frequency. Phenomena such as mode cooling and self-excited oscillations have been shown in systems in which this force is dominant~\cite{Metzger_et_al_08, Metzger&Karral_04, Aubin_et_al_04, Jourdan_et_al_08}. Existing theoretical models that describe these phenomena quantitatively~\cite{Aubin_et_al_04, Marquardt_et_al_06, Paternostro_et_al_06, Ludwig_et_al_07, Liberato_et_al_10} are based on energy or harmonic balance methods which provide good predictions of the system steady state, but lack the ability to fully describe its complex dynamics.

Recently, we have developed a slow envelope dynamical model
of an optomechanical system which includes radiation pressure, thermal force, and changes to mechanical frequency due to absorption heating~\cite{Zaitsev_et_al_11a}. The theoretical predictions, which are derived using a combined harmonic balance and averaging method~\cite{Szemplinska-Stupnicka_book_90}, include all the experimental phenomena shown by optomechanical systems with a bolometric force, such as linear dissipation renormalization and self-excited oscillations. In addition, the model enables prediction of additional nonlinear effects, namely, the change in the sign of the mechanical cubic nonlinear elastic and dissipative terms as function of the optical power incident on the cavity and its exact detuning from resonance~\cite{Aubin_et_al_04}.

Here, we present experimental results that demonstrate all the major dynamical phenomena which are theoretically implied in Ref.~\cite{Zaitsev_et_al_11a}. In order to facilitate the study of optical cavities with micromechanical mirrors spanning a wide range of different geometries and materials, we employ a fiber Bragg grating (FBG) \cite{Snyder&Love_book_83} as a static mirror of the optical cavity. The wave length dependent transparency of the FBG allows us to achieve different coupling conditions between the optical mode inside the cavity and the incident light, thus effectively controlling the cavity's finesse.

A very reasonable fit between theory and experiment is achieved using two distinct geometries of the micromechanical mirror composed of two different metals - AuPd and aluminum. The fits include changes in the linear dissipation, the threshold, the frequency and the amplitude of self-excited oscillations, and the thermally induced frequency shifts under different conditions. In addition, we show optically induced changes in the nonlinear response of the micromechanical mirrors.

\section{Experimental setup}
\label{sec:exp_setup}

\begin{figure}
	\centering
	 \includegraphics[width=3.4in]{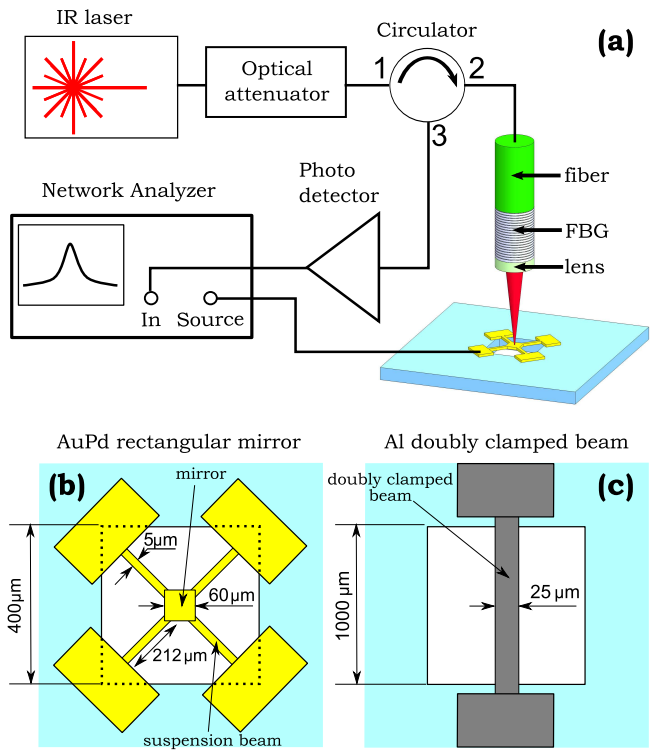}
        \caption{(Color online) \textbf{(a)} The experimental system. In addition to the network analyzer, a number of measuring and excitation schemes can be used, such as homodyne vibration detection with a lock-in amplifier, and direct time sampling of the reflected optical power. A tunable infra-red laser is used, with a maximum output power of $2\milliwatt$. A single mode optical fiber is used to transmit light to and from the sample, allowing reflection measurements (with the aid of a circulator to separate the incident and the reflected beams). A graded index lens at the end of the optical fiber is used to focus the light on the mirror. The focus distance is $\approx 40\micrometer$. A FBG located inside the fiber can serve as a second mirror to create a relatively high-finesse cavity at wavelengths that fall inside the Bragg region. The micromechanical mirror can be capacitively actuated by applying a voltage between the mirror itself and a ground plate located $500\micrometer$ below it (the ground plate is not shown). Panels \textbf{(b)} and \textbf{(c)} exhibit the top views of the micromechanical mirrors employed in the experiments. The thickness of the metal layers is $300\nanometer$ for AuPd samples, and $200\nanometer$ for Al samples.}%
        \label{fig:optomech_exp_setup}
\end{figure}

In the investigated system, an optical resonance cavity is created between a suspended metallic micromechanical mirror, which is free to oscillate in a direction parallel to the optical axis, and a stationary mirror in the form of a FBG as shown in Fig.~\ref{fig:optomech_exp_setup}. The system is located in a vacuum chamber inside a cryostat with a typical pressure of $3\micro\uBar$ and temperature of $77\kelvin$.

A micromechanical mirror is fabricated on a silicon-nitride membrane using electron beam lithography and thermal evaporation of metal. Following these steps, the membrane is removed by electron cyclotron resonance (ECR) plasma etching, and the micromechanical mirror becomes suspended. This fabrication process is similar to the one described in~\cite{Buks&Roukes_01b}.

Two main suspended mirror configurations were used in our experiments, a gold-palladium~($\text{Au}_{0.85}\text{Pd}_{0.15}$) rectangular mirror and an aluminum doubly clamped wide beam. The dimensions of the devices are given in Fig.~\ref{fig:optomech_exp_setup}.

The micromechanical mirror can be actuated capacitively in the direction of the optical axis by applying voltage between the mirror and the ground plate of the package used to mount the sample in the vacuum chamber. The ground plate is parallel to the mirror and located $500\micrometer$ below it.

The FBG, which has a length of $L_B=20\millimeter$, is formed using a phase mask having a period of $1062\nanometer$ on a single mode optical fiber having effective refractive index of $n_{\mathrm{eff}}=1.46$ near the wavelength of $1550\nanometer$. A microlens made from a section of a graded index fiber having length of $0.45$ pitch is spliced to the the end of the fiber \cite{Mao_et_al_07}. A brief analysis of the FBG's optical properties is given in Sec.~\ref{sec:optical_resonance_cavity}.

The optical fiber can be moved in three orthogonal directions by the means of piezomotors with an accuracy of approximately $1\nanometer$. Generally, the fiber is positioned above the center of the micromechanical mirror at a focal distance of the microlens, which is $\approx 40\micrometer$. The length of the optical cavity can be changed by moving the fiber along the optical axis. We control the wavelength and the power of the light incident on the cavity by using a variable wavelength infra-red laser and a variable fiber-optic attenuator, respectively. The light reflected off the cavity back into the fiber is separated by the means of a circulator and converted to an electrical signal by a photodetector. The experimental system is shown in Fig.~\ref{fig:optomech_exp_setup}.

The finesse of the optical cavities created in the presented experiments is of order of ten. We estimate the optical relaxation time to be of order $10^{-12}\sec$. It follows that optical retardation can be neglected in our system, and thus the optical energy stored in the cavity is a function of the momentary displacement of the micromechanical mirror.

\section{Theoretical model}
\label{sec:theor_model}

An extensive theoretical analysis of the dynamics of a micromechanical oscillator acting as a mirror in a low finesse optical cavity based on a slow envelope approximation can be found in Ref.~\cite{Zaitsev_et_al_11a}. Here, we state the main results from that work, and present a short discussion on the FBG optical properties.

	\subsection{Optical cavity}
	\label{sec:optical_resonance_cavity}

The finesse of the optical cavity is limited by loss mechanisms that give rise to optical energy leaking out of the cavity. The main escape routes are through the FBG, through absorption by the metallic mirror, and through radiation, and the corresponding transmission probabilities are respectively denoted by $T_B$, $T_A$ and $T_R$. The transmission probability $T_B$ through the FBG is evaluated using
the coupled mode theory \cite{Snyder&Love_book_83, Poladian_96}
\begin{equation}
T_B=\frac{1}{1+\frac{V_B^{2}\sinh^{2}\left(\sqrt{V_B^{2}-d_B^{2}}\right)  }{V_B^{2}-d_B^{2}}} ,%
\end{equation}
where $d_B=\left(  \omega-\omega_B\right) L_B n_\mathrm{eff}/c$ is the normalized detuning factor, $\omega$ and $\omega_B$ are respectively the laser and Bragg angular frequencies, $c$ is light velocity in vacuum, and $V_B$ is the FBG coupling constant.

Let $x-x_{0}$ be the displacement of the mirror relative to the point $x_{0}$, at which the energy stored in the optical cavity in steady state obtains a local maximum. For a fixed $x$ the cavity reflection probability $R_C$, i.e. the ratio between the reflected (outgoing) and injected (incoming) optical powers in the fiber, is given by
\begin{equation}
	R_C=\frac{\left(  \frac{T_B-T_A-T_R}{T_B+T_A+T_R}\right)^{2}+2\left(\frac{L}{\pi\Gamma}\right)^{2}\left(  1-\cos2\pi\frac{x-x_0}{L}\right)}{1+2\left(\frac{L}{\pi\Gamma}\right)^{2}\left(1-\cos2\pi\frac{x-x_0}{L}\right)},
	\label{eq:opt_power_Rc}
\end{equation}
where $L$ is the distance between two successive resonance positions of the micromechanical mirror (i.e., half the wavelength), and 
\begin{equation*}
	\Gamma=(T_B+T_A+T_R)\frac{L}{2\pi}
\end{equation*}
is the full width at half maximum parameter. The effective optical power $I(x)$
impinging on the suspended micromechanical mirror can be expressed as
\begin{equation}
	I(x)=\frac{I_\text{max}\left(\frac{\Gamma}{2}\right)^2}{\frac{L^2}{2\pi^2}\left[1-\cos 2\pi\frac{x-x_0}{L}\right]+\left(\frac{\Gamma}{2}\right)^2},
	\label{eq:opt_power_periodic_Lorentzian}%
\end{equation}
where 
\begin{equation*}
	I_\text{max}=C_\text{re}I_\text{pump},%
\end{equation*}
is the maximum optical power incident on the mirror, $I_\text{pump}$ is the power of the monochromatic laser light incident on the cavity, and
\begin{equation*}
	C_\text{re}=\frac{4T_B}{\left(T_B+T_A+T_R\right) ^{2}} 
\end{equation*}
is the resonant enhancement factor of the intra-cavity power. Note that for the case of critical coupling, i.e., the case where $T_B=T_A+T_R$, $C_\text{re}= {L}/{\pi\Gamma}$.

The optical power $I(x)$ is a periodic function, which can be approximated by a truncated Fourier series
\begin{equation}
	I(x) \approx \sum_{k=-k_\text{max}}^{k_\text{max}} c_k e^{j2\pi k\frac{x}{L}},
	\label{eq:large_osc_I_Fourier_kmax}%
\end{equation}
where $k_\text{max}$ should be of order of the finesse or larger for the truncation error to be negligible. As shown in Ref.~\cite{Zaitsev_et_al_11a}, $c_k=I_\text{max}\chi\alpha^{|k|}e^{-j2\pi k{x_0}/{L}}$, where $\alpha =1+h-\sqrt{(1+h)^2-1}$, $\chi ={h}/{\sqrt{(1+h)^2-1}}$, and $h={\pi^2\Gamma^2}/{2L^2}$.

	\subsection{Equations of motion}
	\label{sec:eq_of_motion}

The micromechanical mirror can be approximately described as a harmonic oscillator with a single degree of freedom $x$ operating near primary resonance, which is subject to several forces arising from coupling to the optical resonance cavity. In general, a stand alone micromechanical resonator can exhibit nonlinear behavior~\cite{Lifshitz&Cross_08, Zaitsev_et_al_11}. In our experiments, however, the contributions of purely mechanical nonlinearities are negligible, as will be shown in Sec.~\ref{sec:nonlin_effects}.

Following Ref.~\cite{Zaitsev_et_al_11a}, we write the equation of motion as
\begin{equation}
	\ddot x+\frac{\omega_0}{Q}\dot x+\omega_m^2x=2f_m\cos(\omega_0+\sigma_0)t+F_\text{rp}(x)+F_\text{th}(x),
	\label{eq:mech_osc_orig}%
\end{equation}
where a dot denotes differentiation with respect to time $t$, $\omega_0$ is the mechanical resonance frequency of the mirror, $Q$ is the mechanical quality factor, $\omega_m$ is the temperature dependent momentary resonance frequency, $f_m$ is the external excitation force, and $\sigma_0$ is a small detuning of the external excitation frequency from $\omega_0$, i.e., $\sigma_0\ll\omega_0$. The forces resulting from coupling to an optical resonance cavity are the radiation pressure $F_\text{rp}$, and $F_\text{th}$, which is a thermal force that appears due to temperature dependent deformation of the micromechanical mirror \cite{Guckel_et_al_92, Fang&Wickert_94, Fang&Wickert_96, Leong_et_al_08}.

In a wide range of micromechanical resonators, internal tension can strongly affect the resonance frequencies \cite{Pandey_et_al_10, Larsen_et_al_11}. Such systems include the doubly clamped beams and rectangular mirrors with four suspension beams used in our experiments. Changes in the temperature of such devices result in thermal expansion or contraction, which in turn cause changes in internal tension. These changes give rise to a strong temperature dependence of the mechanical resonance frequencies, as will be shown in Sec.~\ref{sec:results}. For small temperature changes, the momentary mechanical frequency $\omega_m$ is assumed to be linearly dependent on the temperature:
\begin{equation}
	\omega_m=\omega_0-\beta(T-T_0),
	\label{eq:omega_m}%
\end{equation}
where $\beta$ is a proportionality coefficient, $T$ is the effective temperature of the mechanical oscillator, and $T_0$ is the temperature of the supporting substrate. In our samples, a significant pretension exists due to thermal evaporation process used to deposit the metals during the manufacturing \cite{Zaitsev_et_al_11, Mintz_thesis_09, Pandey_et_al_10}. The pretension is further increased by cooling the samples to $77\kelvin$. It follows, therefore, that $\beta$ is positive in our experiments, i.e., heating of the sample reduces its resonance frequency.

The effective temperature changes can be described by the following equation,
\begin{equation}
	\dot T=-\kappa(T-T_0)+\eta I(x),
	\label{eq:temp_diff_eq}%
\end{equation}
where $\kappa$ is the effective thermal conductance, and $\eta$ is the effective radiation absorption coefficient. The formal solution of Eq.~\eqref{eq:temp_diff_eq} can be shown to be
\begin{equation*}
	T-T_0=\eta\int_0^t I(x) e^{\kappa(\tau-t)}d\tau,
	\label{eq:T-T0_orig}
\end{equation*}
where the initial transient response term $e^{-\kappa t}\left[T(t=0)-T_0\right]$ has been dropped as insignificant to the long timescale dynamics of the system. This integral relation can be further simplified using the slow envelope approximation. The reader is referred to Ref.~\cite{Zaitsev_et_al_11a} for further details.

Finally, we introduce the radiation dependent forces. The radiation pressure force can be expressed as
\begin{equation}
	 F_\text{rp}(x)=\nu I(x),
	\label{eq:F_rp} \\%
\end{equation}
where
\begin{equation*}
	\nu =\frac{2}{mc},%
\end{equation*}
and where $m$ is the effective mass of the micromechanical mirror. Light absorption by the mirror has been neglected. The thermal force $F_\text{th}$ is assumed to be linear in the temperature difference $T-T_0$, i.e.,
\begin{equation}
	F_\text{th}=\theta(T-T_0),
	\label{eq:Fth}%
\end{equation}
where $\theta$ is a coefficient of proportionality.

The numerical values of all the physical constants introduced above will be evaluated in Sec.~\ref{sec:parameters}.

	\subsection{Slow envelope approximation}
	\label{sec:slow_envelope_approx}
Following Ref.~\cite{Zaitsev_et_al_11a}, the dynamics of the micromechanical mirror can be approximated by a harmonic motion with slow varying amplitude and phase, i.e.,
\begin{subequations}
\begin{equation}
	x(t) \approx A_0+A_1\cos\psi,
	\label{eq:large_osc_x}%
\end{equation}
where
\begin{equation}
	\psi=\omega_0 t+\tilde\phi,
	\label{eq:large_osc_psi}
\end{equation}
	\label{eq:large_osc_x_and_phi}%
\end{subequations}
and where $A_1$ and $\tilde\phi$ are the oscillator's amplitude and phase \cite{Nayfeh_Mook_book_95}, respectively, and $A_0$ is the static mirror displacement due to the action of the radiation dependent forces.

By introducing Eqs.~\eqref{eq:large_osc_x_and_phi} into Eq.~\eqref{eq:mech_osc_orig} and  using a combined harmonic balance - averaging method~\cite{Szemplinska-Stupnicka_book_90}, one can derive the following relations which describe the slow envelope behavior of the mirror [see Eqs.~(22)-(25) in Ref.~\cite{Zaitsev_et_al_11a}]:
\begin{subequations}
\begin{equation}
	A_0 \approx \frac{1}{\Omega^2}\left[2P_1\beta\eta \frac{\omega_0\kappa}{\kappa^2+\omega_0^2}A_1+P_0 \left(\nu+\frac{\theta\eta}{\kappa}\right)\right],
	\label{eq:large_osc_A0}%
\end{equation}
\begin{multline}
	\dot A_1 = -\left(\frac{\omega_0}{2Q}+2P_2\beta\eta\frac{\omega_0}{\kappa^2+4\omega_0^2}\right)A_1\\
-P_1\eta\frac{\omega_0}{\kappa^2+\omega_0^2}\left(2\beta A_0+\frac{\theta}{\omega_0}\right)\\
-\frac{f_m}{\omega_0} \sin\phi,
	\label{eq:large_osc_polar_evolution_dotA1}%
\end{multline}
and
\begin{multline}
	A_1\dot\phi =-\left(\sigma_0+\Delta\omega_0+P_2\beta\eta \frac{\kappa}{\kappa^2+4\omega_0^2}\right)A_1\\
-P_1\eta\frac{\kappa}{\kappa^2+\omega_0^2}\left(2\beta A_0+\frac{\theta}{\omega_0}\right)-P_1\frac{\nu}{\omega_0}\\
-\frac{f_m}{\omega_0} \cos\phi,
	\label{eq:large_osc_polar_evolution_phi}
\end{multline}
	\label{eq:large_osc_polar_evolution_eq}
\end{subequations}
where $\phi=\tilde\phi-\sigma_0 t$ (the detuning $\sigma_0$ is assumed small),
\begin{equation}
	\Omega=\omega_0-\frac{\beta\eta}{\kappa}P_0=\omega_0-\Delta\omega_0,
	\label{eq:large_osc_Omega}%
\end{equation}
and
\begin{equation*}
	P_n(A_0,A_1)=\sum_{k=-k_\text{max}}^{k_\text{max}} j^n c_k e^{j2\pi k\frac{A_0}{L}} J_n\left(2\pi k\frac{A_1}{L}\right),
	\label{eq:large_osc_Pn}%
\end{equation*}
where $J_n(z)$ is the Bessel function of order $n$.

The term $\Delta\omega_0$ represents a small mechanical frequency correction due to the averaged heating of the micromechanical mirror vibrating with an amplitude $A_1$. As will be shown in Sec.~\ref{sec:results}, this correction accounts for the dominant part of the resonance frequency shift measured in our experiments.

	\subsection{Small amplitude oscillations}
	\label{sec:small_osc}
	
The evolution equations~\eqref{eq:large_osc_polar_evolution_eq} can be conveniently simplified if the vibration amplitude of the micromechanical mirror is small compared to the optical resonance width parameter $\Gamma$. In this case, 
\begin{equation*}	
	A_{0s}= \frac{I_0}{\Omega_s^2}\left(\nu+\frac{\theta\eta}{\kappa}\right), \\ \label{eq:small_osc_A0s}%
\end{equation*}
and
\begin{subequations}
\begin{align}
	\dot A_{1s} & =-\gamma A_{1s}-\frac{r}{4}A_{1s}^3-\frac{f_m}{\omega_0} \sin\phi, \\
	A_{1s}\dot\phi & =-\left(\sigma_0+\Delta\omega_s\right)A_{1s}+\frac{q}{4}A_{1s}^3-\frac{f_m}{\omega_0} \cos\phi,
\end{align}
	\label{eq:small_osc_polar_evolution_eq_qr}%
\end{subequations}
where the subscript 's' denotes small amplitude oscillations, $I_0=I(x=0)$, and where
\begin{subequations}
\begin{align}
	& \Delta\omega_{0s} =\frac{\beta\eta}{\kappa}I_0, \\
	& \Omega_s =\omega_0-\Delta\omega_{0s},
\end{align}
\begin{equation}
	\gamma =\frac{\omega_0}{2Q}+\eta\frac{\omega_0}{\kappa^2+\omega_0^2}\left(\beta A_{0s}+\frac{\theta}{2\omega_0}\right)I_0',
	\label{eq:small_osc_gamma}
\end{equation}
\begin{multline}
	\Delta\omega_s =\Delta\omega_{0s}\\
	+\left[\frac{\nu}{2\omega_0}+\frac{\eta\kappa}{\kappa^2+\omega_0^2}\left(\beta A_{0s}+\frac{\theta}{2\omega_0}\right)\right]I_0', \label{eq:small_osc_omegas}
\end{multline}
	\label{eq:small_osc_params}%
\end{subequations}
and
\begin{subequations}
\begin{align}
	q & =-\frac{\beta\eta}{2\kappa}\frac{3\kappa^2+8\omega_0^2}{\kappa^2+4\omega_0^2}I_0'',
	\label{eq:small_osc_q} \\
	r & =\beta\eta\frac{\omega_0}{\kappa^2+4\omega_0^2}I_0''.
	\label{eq:small_osc_r}
\end{align}
	\label{eq:small_osc_q_and_r}%
\end{subequations}
Note that a prime denotes differentiation with respect to $x$, i.e., $I_0'=dI(x=0)/dx$.

The evolution equations~\eqref{eq:small_osc_polar_evolution_eq_qr} describe a Duffing-like nonlinear oscillator with nonlinear damping~\cite{Nayfeh_Mook_book_95, Zaitsev_et_al_11}. Interestingly enough, the sign of the nonlinearities depends on the sign of the second derivative of $I_0$ with respect to $x$. For example, Eq.~\eqref{eq:small_osc_q} predicts that the system should exhibit hardening behavior near the maximum of the optical resonance (more precisely, in the region where $I_0''<0$, i.e., $|x_0|<\Gamma/2\sqrt{3}$), and softening behavior otherwise. This effect is experimentally illustrated in Sec.~\ref{sec:results} for both types of micromechanical mirrors studied.

Another interesting effect that depends on the optical detuning $x_0$ of the micromechanical mirror is the change in the effective linear damping coefficient $\gamma$ as function of $I_0'$, which is evident from Eq.~\eqref{eq:small_osc_gamma}. From the experimental point of view, it is convenient to introduce the effective quality factor as
\begin{equation}
	\frac{1}{Q_\text{eff}}=\frac{2\gamma}{\Omega_s}.
	\label{eq:Qeff}
\end{equation}

We expect an increase in $1/Q_\text{eff}$ as compared to the purely mechanical value $1/Q$ in the region in which $I_0'>0$ ($x_0>0$), corresponding to the mode "cooling" effect \cite{Braginsky_et_al_70, Kimble_et_al_01, Genes_et_al_08, Restrepo_et_al_10}, and, conversely, decrease in the effective dissipation for the values of $x_0$ at which $I_0'<0$, i.e., $x_0<0$. In this region, the effective dissipation may become arbitrarily small and even change sign, resulting in a Hopf bifurcation followed by possible self-excited limit cycle oscillations~\cite{Strogatz_book_94, Zaitsev_et_al_11a}.

	\subsection{Self-excited oscillations}
	\label{sec:self_osc}

Self-excited oscillations may occur in a system described by Eqs.~\eqref{eq:large_osc_polar_evolution_eq} if a stable limit cycle \cite{Arnold_book_88, Strogatz_book_94} exists in absence of external excitation. In other words, a nonzero solution of the following equation, together with Eq.~\eqref{eq:large_osc_A0}, is required:
\begin{multline}
	 \left(\frac{\omega_0}{2Q}+2P_2\beta\eta\frac{\omega_0}{\kappa^2+4\omega_0^2}\right)A_1\\
	+P_1\eta\frac{\omega_0}{\kappa^2+\omega_0^2}\left(2\beta A_0+\frac{\theta}{\omega_0}\right)=0.
	\label{eq:self_osc_dotA1}
\end{multline}
Again, we refer the reader to Ref.~\cite{Zaitsev_et_al_11a} for a full analysis of different bifurcations and limit cycles types which may appear in systems under study. In our experiments, a single stable limit cycle is observed, appearing beyond the threshold of a supercritical Hopf bifurcation. As expected, the region in which the system develops self-excited oscillations coincides with the region of negative effective linear dissipation, i.e., $\gamma<0$ [see Eq.~\eqref{eq:small_osc_gamma}], in which the zero amplitude solution $A_1=0$ is unstable.

For a given nonzero oscillation amplitude $A_1$, the oscillation frequency correction is given by Eq.~\eqref{eq:large_osc_polar_evolution_phi}:
\begin{multline}
	\phi=\Bigg\{-\Delta\omega_0-P_2\beta\eta \frac{\kappa}{\kappa^2+4\omega_0^2}\\
-\frac{P_1}{A_1}\left[\eta\frac{\kappa}{\kappa^2+\omega_0^2}\left(2\beta A_0+\frac{\theta}{\omega_0}\right)+\frac{\nu}{\omega_0}\right]\Bigg\}t\\
	=-\Delta\omega t.
	\label{eq:self_osc_phi}%
\end{multline}%

\section{Parameter evaluation}
\label{sec:parameters}

We now turn to evaluate the physical parameters, which are defined in the previous section, for a rectangular $\text{Au}_{0.85}\text{Pd}_{0.15}$ mirror, whose dimensions are given in Fig.~\ref{fig:optomech_exp_setup}.

The environment temperature in our experiments is $T_0=77\kelvin$. Using a weighted averaging of the values for gold~\cite{Geballe&Giauque_52, White_53} and palladium~\cite{Veal&Rayne_64}, we estimate the values of the density $\rho=18.2\times 10^3 \kg\meter^{-3}$, the mass-specific heat capacity $C_m=103 \joule\kg^{-1}\kelvin^{-1}$, and the thermal conductivity $k=281\watt\meter^{-1}\kelvin^{-1}$ for $\text{Au}_{0.85}\text{Pd}_{0.15}$ mirror at $77\kelvin$. Although by no means precise, this simple averaging method provides a reasonable accuracy in our case.


We take the effective mass of the micromechanical resonator to be the mass of the mirror (the mass of the suspension beams is neglected). Using the mirror's dimensions and the density value derived above, we find that $m \approx 20\nanogram$.

The effective thermal relaxation rate $\kappa$ can be evaluated as follows:
\begin{equation*}
	\kappa=4\frac{300\nanometer\times 5\micrometer}{212\micrometer}\frac{k}{m C_\text{m}}=3.9\times 10^3\frac{1}{\sec}.
\end{equation*}

In order to estimate the value of the effective radiation absorption coefficient $\eta$, the reflectivity of the micromechanical mirror must be known. In the literature \cite{Bennett&Ashley_65, Yu&Spicer_68}, experimental values between 98\% and 99\% are given. We find that an empirical value of 98.4\% fits our experimental results. It follows that
\begin{equation*}
	\eta=\frac{1-0.984}{mC_\text{m}}=7.9\times 10^6 \frac{\kelvin}{\joule}.
\end{equation*}

The high reflectivity of the micromechanical mirror allows us to neglect any absorption when estimating the radiation pressure coefficient, resulting in:
\begin{equation*}
	\nu=\frac{2}{mc}=339 \frac{\newton}{\kg\watt}.
\end{equation*}

The estimation of the thermal frequency shift coefficient $\beta$ is not straightforward. The order of magnitude can be estimated by measuring the mechanical resonance frequency of the mirror at room temperature and at $77\kelvin$ ($106\kilohertz$ and $160\kilohertz$, respectively), resulting in $\beta\approx 0.012\omega_0/1\kelvin$. However, in order to give an accurate estimation of $\beta$ for small temperature changes around $77\kelvin$ (or any other ambient temperature), one would require preexisting knowledge of the tension inside the sample, the exact relation between the tension and the mechanical resonance frequency, and, most importantly, the exact temperature distribution inside the sample due to nonuniform heating by a focused laser beam. This data is not readily available from our measurements. Therefore, we treat $\beta$ as one of the fitting parameters. The best fit is achieved for
\begin{equation*}
	\beta=\frac{0.01\omega_0}{1\kelvin}=1.006\times 10^4 \frac{\radian}{\sec\kelvin},
\end{equation*}
which is remarkably similar to the value estimated above.

Although the majority of the parameters defined in Sec.~\ref{sec:theor_model} can be evaluated using general physical considerations or direct measurements, $\theta$ is not easily determined, because the physical processes responsible for the appearance of the thermal force $F_\text{th}$ are not well identified. Therefore, we derive the value of $\theta$ from experiment. The best fit is achieved when $\theta$ is taken to be
\begin{equation*}
	\theta=740 \frac{\newton}{\kg\kelvin}.
\end{equation*}

By estimating the ratio
\begin{equation*}
	\frac{\nicefrac{\theta\eta}{\kappa}}{\nu} \sim 10^3,
\end{equation*}
it follows from Eqs.~\eqref{eq:large_osc_polar_evolution_eq} that the radiation pressure effects in our system are negligible compared to the effects of the thermal force $F_\text{th}$.

\section{Results and Discussion}
\label{sec:results}

Here, we present a comparison between the experimental behavior of three different micromechanical mirrors and the theoretical predictions given in Sec.~\ref{sec:theor_model}. For convenience, the main mechanical properties of these mirrors are summarized in Table~\ref{table:samples_properties}.

\begin{table}[ht]
	\centering
	\renewcommand{\arraystretch}{2}
	\begin{tabular}{ c | c | c }
		\hline
		sample & parameter & value \\[0.5ex]
		\hline \hline
		Sample I & $\omega_0$ & $2\pi\times 160.088 \kilohertz$\\[0.5ex]
		AuPd mirror &   $Q$ & $2.43\times 10^5$ \\[0.5ex]
		\hline
		Sample II & $\omega_0$ & $2\pi\times 148.495 \kilohertz$\\[0.5ex]
		AuPd mirror &   $Q$ & $6.75\times 10^4$ \\[0.5ex]
		\hline
		Sample III & $\omega_0$ & $2\pi\times 61.25 \kilohertz$\\[0.5ex]
		Al beam &   $Q$ & $1\times 10^5$ \\[0.5ex]
		\hline
	\end{tabular}
	\caption{Mechanical properties of the samples used in this study.}
	\label{table:samples_properties}
\end{table}
	\subsection{Optical resonance cavity}
	\label{sec:res_opt_cavity}

In our experiments, we tune the optical wavelength to a value at which the reflection from the cavity becomes virtually zero at the resonance, a condition known as critical coupling. In general, this critical coupling wavelength is at the edge of the Bragg region, where the FBG reflectivity changes from almost zero to almost unity. For example, the optical wavelength used for measurements of square AuPd micromechanical mirrors is $1548.83\nanometer$. It follows that the distance between the subsequent minimums in the reflection $R_C$ (or, conversely, peaks in $I(x)$) is $L=774.4\nanometer$, allowing us to calibrate the vertical displacement of the fiber at any temperature. A typical finesse of the cavity is between 6 and 11, i.e., $70\nanometer<\Gamma<140\nanometer$. In general, each time the cavity is optically tuned by realignment of the fiber, a slightly different finesse can be expected, due to inaccuracies in the fiber positioning.

\begin{figure}
	\centering
	 \includegraphics[width=3.4in]{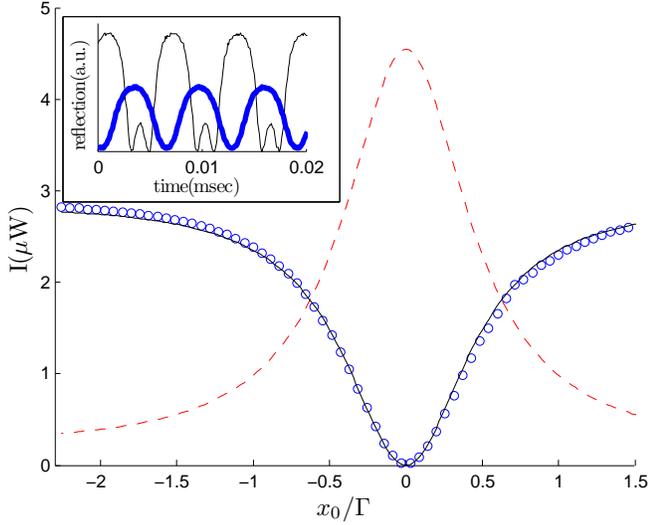}
        \caption{(Color online) Optical reflection in the vicinity of an optical resonance. The incident optical power is $I_\text{pump}=3\microwatt$. Blue circles denote the measured reflected power. Solid black line is a theoretical reflection fit [see Eq.~\eqref{eq:opt_power_Rc}]. Dashed red line represents the theoretical values of the optical power $I_0$ incident on the micromechanical mirror [see Eq.~\eqref{eq:opt_power_periodic_Lorentzian}]. In this case, $\Gamma=134.5\nanometer$, and the finesse is 5.8. In the inset, two time domain traces of the photodetector output signal are presented, corresponding to a steady oscillation of the micromechanical mirror with a large (thin black line) and a small (bold blue line) amplitude.}
        \label{fig:opt_refl_example}
\end{figure}

Instantaneous changes in the micromechanical beam displacement $x$ cause changes in the reflected power according to Eq.~\eqref{eq:opt_power_Rc}. The signal at the output of the photodetector can be translated into actual displacement values using the calibration discussed above. An example of the reflected optical power vs.\ the optical cavity detuning $x_0$, together with sample time traces of mirror oscillatory movement is shown in Fig.~\ref{fig:opt_refl_example}. It is evident from this figure that the theory presented in Sec.~\ref{sec:optical_resonance_cavity} provides a good analytical description of the experimental measurements of the optical cavity behavior.

	\subsection{Linear damping}
	\label{sec:lin_damping}
	
We begin our experimental study with investigation of what is arguably the most important prediction of the theoretical model - the possibility of a significant change in the effective dissipation in the vicinity of an optical resonance. In order to measure the effective quality factor $Q_\text{eff}$ defined in Eq.~\eqref{eq:Qeff} at different optical powers $I_\text{pump}$ and cavity detunings $x_0$, we capacitively excite the micromechanical mirror at its apparent resonance frequency for a short period of time, and then allow the system to decay freely to the zero amplitude steady state. During this free ring down process, the slow envelope of the mechanical oscillations is measured by the means of a lock-in amplifier. The resulting slow envelope is fitted to an exponential decay function proportional to $e^{-2\gamma t}$, providing an estimate of the linear dissipation constant. It is important to keep the vibration amplitude small compared to $\Gamma$, so the nonlinearities introduced by the detection system and the optomechanical coupling [see Eqs.~\eqref{eq:small_osc_q_and_r}] remain negligible.

\begin{figure}
	\centering
	 \includegraphics[width=3.4in]{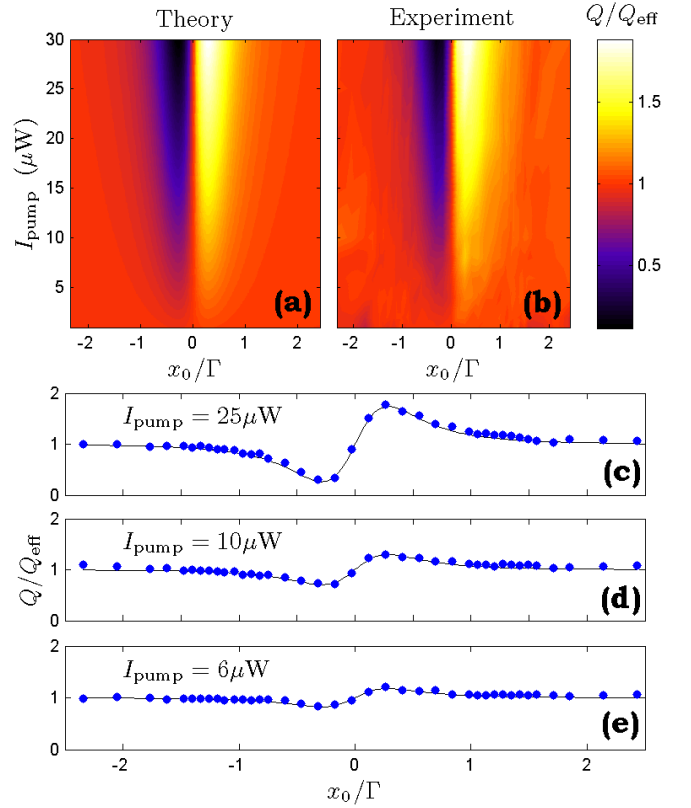}
        \caption{(Color online) Changes in the effective quality factor $Q_\text{eff}$ [see Eq.~\eqref{eq:Qeff}] as function of the optical detuning $x_0$ and the optical power incident on the cavity, $I_\text{pump}$. Panel \textbf{(a)} shows the theoretical value of $Q/Q_\text{eff}$, whereas panel \textbf{(b)} shows the measured data. In the case of a rectangular AuPd mirror presented here, $\omega_0=2\pi\times 160.088 \kilohertz$, $Q=2.43\times 10^5$, and the cavity finesse is 7.3. Panels \textbf{(c)}, \textbf{(d)} and \textbf{(e)} show cross sections of the top color maps at different values of $I_\text{pump}$. Blue dots represent the experimental values of $Q/Q_\text{eff}$, while solid black lines represent the theoretical results. The values of all system parameters used in the fit are similar to those given in Sec.~\ref{sec:parameters}. The temperature is $77\kelvin$.}
        \label{fig:optomech_exp_Qeff}
\end{figure}

The results presented in Fig.~\ref{fig:optomech_exp_Qeff} show a good match between the experimental values of $Q_\text{eff}$ and the theoretical predictions. The measurement was done at $77\kelvin$ using Sample I (see Table~\ref{table:samples_properties}). The values of all the system parameters used in the fit are similar to those given in Sec.~\ref{sec:parameters}.

	\subsection{Nonlinear stiffness and damping effects}
	\label{sec:nonlin_effects}
	
\begin{figure}
	\centering
	 \includegraphics[width=3.4in]{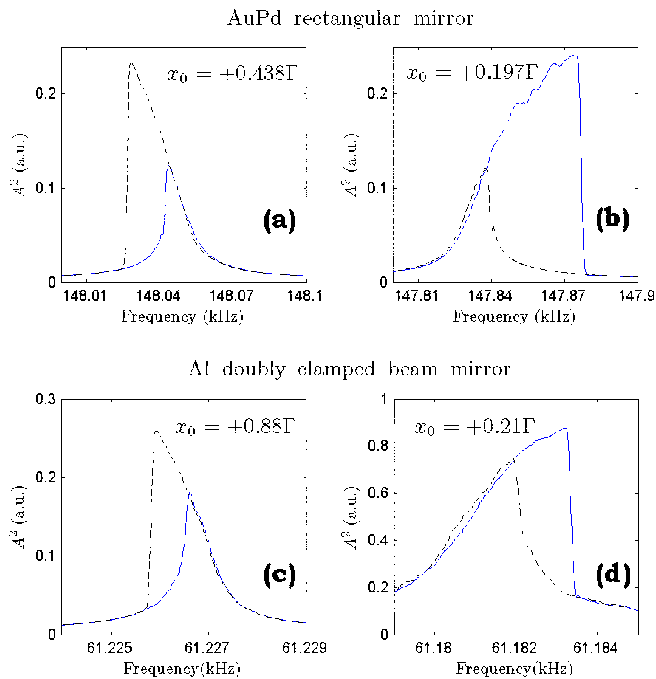}
        \caption{(Color online) Changes in the small amplitude nonlinear behavior as a function of the optical detuning $x_0$. It follows from Eqs.~\eqref{eq:small_osc_q_and_r} that the micromechanical mirror should exhibit hardening ($q>0$) behavior \cite{Nayfeh_Mook_book_95} in the region $|x_0|<\Gamma/2\sqrt{3}=0.289\Gamma$, and softening behavior outside this region. This effect is illustrated for two different samples. \textbf{(a)}, \textbf{(b)} AuPd rectangular mirror measurements. \textbf{(c)}, \textbf{(d)} Al doubly clamped beam mirror measurements. In both cases, the micromechanical mirrors are excited capacitively and the frequency response is measured when sweeping the excitation frequency up (solid blue line) and down (dashed black line). As expected, the frequency responses \textbf{(a)} and \textbf{(c)}, for which $x_0>\Gamma/2\sqrt{3}$, exhibit softening elastic nonlinearity, while the frequency responses \textbf{(b)} and \textbf{(d)}, corresponding to $x_0<\Gamma/2\sqrt{3}$, show hardening. The optical power incident on the cavity is $14\microwatt$ for AuPd rectangular mirror and $2.5\microwatt$ for Al doubly clamped beam mirror.}
        \label{fig:optomech_exp_nonlin}
\end{figure}

The nonlinear effects described in Eqs.~\eqref{eq:small_osc_q_and_r} have been observed in all our samples. Here, we present the small amplitude frequency response of Samples II and III (see Table~\ref{table:samples_properties}), taken at different cavity detuning values. It follows from the discussion in Sec.~\ref{sec:small_osc} that the elastic nonlinearity coefficient $q$ should change sign at $x_0=\pm\Gamma/2\sqrt{3}=\pm0.289\Gamma$. Outside this region the system is expected to behave as a softening Duffing-like oscillator, while in the region around the optical resonance the behavior should be hardening. The experimental results presented in Fig.~\ref{fig:optomech_exp_nonlin} confirm this prediction qualitatively.

In general, nonlinear elastic and dissipative effects in micromechanical systems can have a non negligible impact on the dynamics of these systems \cite{Nayfeh_Mook_book_95, Lifshitz&Cross_08, Almog_et_al_06a, Almog_et_al_06b, Zaitsev_et_al_11}. In the theoretical treatment in Ref.~\cite{Zaitsev_et_al_11a}, the nonlinear effects which do not stem from optomechanical coupling are described by the cubic nonlinearity coefficients $\alpha_3$ and $\gamma_3$ [see Eq.~(7) in Ref.~\cite{Zaitsev_et_al_11a}]. However, the experimental results show that in our samples the nonlinearities introduced by the optomechanical coupling are much stronger than any preexisting nonlinear effects, at least at relatively high optical powers. Therefore, in the present work, we have neglected all nonlinearities that do not arise due to the interaction with the optical system, i.e., $\alpha_3=0$ and $\gamma_3=0$.

	\subsection{Self-excited oscillations}
	\label{sec:res_self_osc}

\begin{figure}
	\centering
	 \includegraphics[width=3.4in]{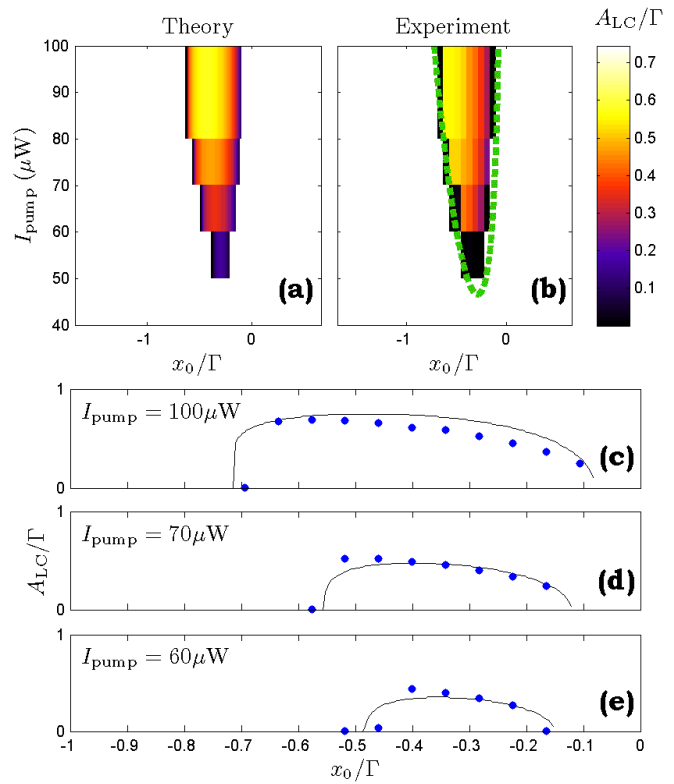}
        \caption{(Color online) The self oscillation amplitude $A_\text{LC}$ as function of the optical detuning $x_0$ and optical power incident on the cavity, $I_\text{pump}$. Panel \textbf{(a)} shows theoretical fit, whereas panel \textbf{(b)} shows measured data. The color represents the value $A_\text{LC}/\Gamma$. The dashed bold green line in the experimental color map \textbf{(b)} represents the theoretical threshold of self-oscillations. In the case of a rectangular AuPd mirror presented here, $\omega_0=2\pi\times 160.088 \kilohertz$, $Q=2.43\times 10^5$, and the cavity finesse is 5.9. Panels \textbf{(c)}, \textbf{(d)} and \textbf{(e)} show cross sections of the top color maps at different values of $I_\text{pump}$. Blue dots represent the experimental values of $A_\text{LC}/\Gamma$, while solid black lines represent the theoretical results. The values of all the system parameters used in the fit are similar to those given in Sec.~\ref{sec:parameters}. The temperature is $77\kelvin$.}%
        \label{fig:optomech_exp_Alc}%
\end{figure}

All the samples used in our experiments exhibit the phenomenon of self sustained oscillations (i.e., stable limit cycle) above certain threshold of the incident optical power. As expected, these self oscillations always occur when $x_0<0$, i.e., in the region in which $I_0'<0$. The onset of the self oscillation can be predicted by calculating the effective linear dissipation coefficient $\gamma$ given in Eq.~\eqref{eq:small_osc_gamma}. Self oscillations occur when $\gamma$ becomes negative. The amplitude and the frequency of the stable limit cycle can be found by solving Eqs.~\eqref{eq:self_osc_dotA1} and~\eqref{eq:self_osc_phi}, respectively. A comparison between the experimentally measured self oscillation amplitudes of Sample I (see Table~\ref{table:samples_properties}) and the corresponding solutions of Eq.~~\eqref{eq:self_osc_dotA1} is shown in Fig.~\ref{fig:optomech_exp_Alc}.

It should be emphasized that the theoretical predictions presented in Figs.~\ref{fig:optomech_exp_Qeff} and~\ref{fig:optomech_exp_Alc} are both based on the same set of physical parameters presented in Sec.~\ref{sec:parameters} and on the mechanical properties of Sample I, given in Table~\ref{table:samples_properties}, and differ only in the value of $\Gamma$, which changes between different experiments, as explained above. It follows, therefore, that the theoretical model presented here can successfully describe both small vibration behavior and self oscillations with large amplitudes. The parameters extracted from experiments in one of these two modes of operation can be used to predict the dynamics of the system in the other mode.

While the theoretical fit shown in Fig.~\ref{fig:optomech_exp_Alc} is very reasonable, a hysteresis phenomenon exists in the experimental system which can not be explained by the model described above. The data presented in Fig.~\ref{fig:optomech_exp_Alc} was taken while sweeping the optical power from low to high values for a fixed value of $x_0$. However, when the optical power is swept in the opposite direction, i.e., from high to low, the self oscillations disappear at lower values of $I_\text{max}$. The difference in the threshold optical power can be as large as 30\%. It should be mentioned that a theoretical analysis of this specific system with parameters derived in Sec.~\ref{sec:parameters} does not predict other stable limit cycles, although multiple stable limit cycles~\cite{Hane&Suzuki_96, Aubin_et_al_04, Metzger_et_al_08}, as well as subcritical Hopf bifurcations~\cite{Zaitsev_et_al_11a} are possible in systems of this type. In the system considered, the hysteresis can be possibly attributed to changes in the heating pattern and the temperature distribution in the vibrating mirror, which cannot be captured by a model with a single degree of freedom used in our analysis. A multi-mode continuum mechanics analysis of the investigated optomechanical system may provide additional insight into this hysteresis phenomenon.

It remains to determine whether the theoretical frequency shift correctly predicts the corresponding experimental results both in the case of small vibrations [see Eq.~\eqref{eq:small_osc_omegas}] and in the case of self-excited oscillations [see Eq.~\eqref{eq:self_osc_phi}]. To this end, we employ Sample II (see Table~\ref{table:samples_properties}) and measure the spectral power density of the reflected light at the vicinity of the sample's mechanical resonance frequency $148.495 \kilohertz$. In this experiment, the sample is not excited externally. The incident optical power is tuned so the system is expected to develop self oscillations at some region of negative optical detunings $x_0$. For other values of $x_0$, thermal vibrations manifest themselves as a thermal peak, whose frequency is shifted by $-\Delta\omega_s$ from $\omega_0$. By taking the spectrum traces at different values of $x_0$, we are able to measure both the frequency of the small oscillations (i.e., the frequency of the thermal peak) and the self oscillation frequency.

\begin{figure}
	\centering
	 \includegraphics[width=3.4in]{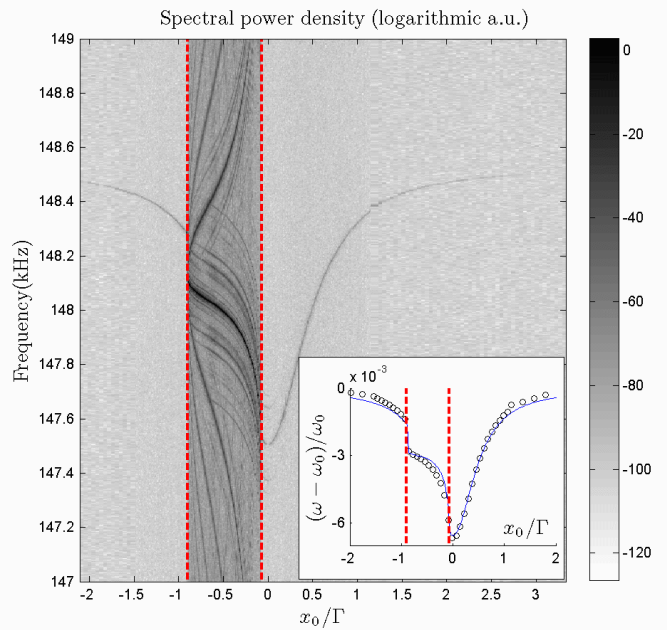}
        \caption{(Color online) Spectral power density of the reflected optical power as function of frequency and optical detuning $x_0$. The mechanical resonance frequency is $148.495 \kilohertz$. The incident optical power is $I_\text{pump}=140\microwatt$. The color represents the spectral power density in arbitrary logarithmic units. The region in which self oscillations occur is denoted by dashed red lines. The thermal motion peak can be readily recognized outside this region. The abundance of additional peaks in the self-excited oscillations domain may be attributed to the nonlinearity of the detection system
and to mixing between higher mechanical modes.
In the inset, the experimental values of frequency shift (blue circles) both for thermal peak and for self oscillations are plotted together with the theoretical predictions (solid black line) given by Eqs.~\eqref{eq:small_osc_omegas} and~\eqref{eq:self_osc_phi}.}
        \label{fig:optomech_exp_spectrum}
\end{figure}

The experimental results together with a theoretical frequency shift fit are presented in Fig.~\ref{fig:optomech_exp_spectrum}. A very reasonable fit between theory and experiment is seen in this figure. Interestingly enough, the thermal frequency correction $\Delta\omega_0$ [see Eq.~\eqref{eq:large_osc_Omega}] constitutes at least 98\% of the frequency shift in the entire measured region.

\section{Summary}
\label{sec:summary}

In this work, we experimentally investigate the dynamics of a metallic micromechanical mirror which is one of the two mirrors that form an optical resonance cavity. The other, static mirror is implemented as a fiber Bragg grating. This unique design allows one to tune the optical cavity operating conditions to the critical coupling domain simply by controlling the wavelength of the incident light.

The finesse of our experimental optical cavities is of order ten. Therefore, all optical retardation effects can be neglected, and only thermal retardation can play a significant role in the dynamics of the micromechanical mirror. A theoretical model describing such a system was developed in Ref.~\cite{Zaitsev_et_al_11a}. Here, the main results are stated, both for small amplitude forced oscillations and for self sustained oscillations.

Theory predicts that coupling of the micromechanical oscillator to an optical cavity will result in changes in its effective linear dissipation, nonlinear elastic and dissipation constants, and the mechanical resonance frequency. Stable limit cycles (i.e., self sustained oscillations) will occur if the effective linear dissipation becomes negative. In addition, multiple limit cycles may be present under certain conditions. Two main optomechanical coupling mechanisms are postulated, both intermediated by heating. The first is mechanical frequency change due to heating, the other is a direct force which is a function of the temperature difference between the mirror and the environment (thermal force). The radiation pressure force is shown to be negligible in our experiments.

In the present work all the theoretical predictions mentioned above are validated by the means of micromechanical mirrors with two very different geometries (rectangular mirror with four orthogonal suspensions and a wide doubly clamped beam). The majority of the physical parameters are derived either from general considerations or independent measurements. A very reasonable quantitative agreement between the linear dissipation changes, the self oscillation amplitudes, and the frequency shifts are achieved. In addition, the theoretically predicted changes in nonlinear behavior are demonstrated for both mirror configurations.

Despite the general success of the theoretical fits of the experimental data, it is evident that a simple single degree of freedom model cannot explain some of the observed phenomena, most importantly the exact process that gives rise to the thermal force. Another unexplained phenomenon is the optical power threshold hysteresis occurring in the self oscillation measurements. Both effects can be possibly attributed to localized changes in heating and temperature distribution, and continuum mechanics approach is required in order to model them correctly.

\section*{Acknowledgments}
We would like to thank O. Gottlieb for many fruitful discussions and important comments. This work is supported by the German Israel Foundation under grant 1-2038.1114.07, the Israel Science Foundation under grant 1380021, the Deborah Foundation, Eliyahu Pen Research Fund, Russell Berrie Nanotechnology Institute, the European STREP QNEMS Project and MAFAT.

\bibliographystyle{unsrtnat}
\bibliography{optomech_exp_arxiv.bbl}

\end{document}